# An account of Natural material based Non Volatile Memory Device

Farhana Yasmin Rahman, Debajyoti Bhattacharjee and Syed Arshad Hussain*

Thin Film and Nanoscience Laboratory, Department of Physics, Suryamaninagar, Agartala 799022, West Tripura, Tripura, India

**Farhana Yasmin Rahman**
farhana.physics@tripurauniv.ac.in

**Debajyoti Bhattacharjee**
debu_bhat@hotmail.com

**Syed Arshad Hussain**
* Corresponding author
Email: sa_h153@hotmail.com, sahussain@tripurauniv.ac.in (SAH)
Ph: +919402122510 (M), Fax: +913812374802 (O)
ORCID: 0000-0002-3298-6260

**Abstract:**

The development in electronic sector has brought a remarkable change in the life style of mankind. At the same time this technological advancement results adverse effect on environment due to the use of toxic and non degradable materials in various electronic devices. With the emergence of environmental problems, the green, reprogrammable, biodegradable, sustainable and environmental-friendly electronic devices have become one of the best solutions for protecting our environment from hazardous materials without compromising the growth of the electronic industry. Natural material has emerged as the promising candidate for the next generation electronic devices due to its easy processing, transparency, flexibility, abundant resources, sustainability, recyclability, and simple extraction. This review targets the characteristics, advancements, role, limitations, and prospects of using natural materials as the functional layer of a resistive switching memory device with a primary focus on the switching/memory properties. Among the available memory devices, resistive random access memory (RRAM), write once read many (WORM) unipolar memory etc. devices have a huge potential to become the non-volatile memory of the next generation owing to their simple structure, high scalability, and low power consumption. The motivation behind this work is to promote the use of natural materials in electronic devices and attract researchers towards a green solution of hazardous problems associated with the electronic devices.

**Introduction:**

Development of science and technology, particularly opto-electronics has an enormous impact on human society, making our day to day life easier, more comfortable and convenient. At the same time such technological development has tremendous negative impact on our environment like air pollution, climate change, decline in natural material resources, technological inequality etc. Modern electronics and optoelectronics are mainly based on silicon and they have very short life span. For example, the average life of smart phone and laptops are less than four years[1] Some sensing devices, electronic toys etc. have even shorter life spans and generate e-waste. Disposal of electronic gadgets has become a major concern.

United Nations (UN) in 2015 has adopted the 2030 agenda for sustainable development and suggested 17 sustainable development goals[2]. According to the $12^{th}$ goal, the electronic waste has grown by 38% and less than 20% of that waste has been recycled. According to report the volume of e-waste in India will reach nearly 2MT by 2025[3].

Memory or data storage device has contributed a large share to this e-waste. This is because in the present big data era the global data storage market size is projected to grow from USD 13.6 billion in 2022 to USD 38.5 billion by 2027 at a compound annual growth rate of 23%[4]. Accordingly, the requirement of data storage device has grown enormously, therefore the biggest challenge to the scientists and engineers is to find out alternative technology which is sustainable to the environment.

Materials play a big role towards the development of sustainable electronic devices as the degradability or compatibility of the device with environment depends on the materials. More than 90% of present day electronics devices are built using Si, which has many disadvantages in terms of degradability, scalability as well as abundance.

So one of the main goal of the researchers to find alternative materials suitable for electronics including memory element[5]. Several materials of biological origin have already been investigated with the aim of achieving sustainable technology[6]. In this regard, the concept of "all natural electronic devices" has come up by utilizing natural materials in opto-electronic applications[7].

On the other hand, resistive switching is one of the emerging phenomenon for the realization of next generation nonvolatile memory. Resistive switching (RS) devices generally comprise of a two terminal geometry in which the active material is sandwiched between two conductive electrodes like metal-insulator-metal (MIM) structure[8–24]. In this device after application of suitable bias, resistances can be toggled in between two definite states as high resistance state (HRS) and low resistance state (LRS)[25–29]. In such devices most of the cases, changes in resistance states are considered as information in terms of high and low resistances. Based on the switching between resistance or nature of operation, these devices are suitable for applications like WORM memory, RRAM etc.[21, 30–36]. Comparison of performance / memory parameters of RRAM with the current complementary metal oxide semiconductor (CMOS) / magnetic memory technologies has been given in table 1.

Table 1: Performance comparison for RRAM, CMOS and Magnetic memory

| Parameter | Resistive Switching memory | CMOS memory | | Magnetic memory | Reference |
|---|---|---|---|---|---|
| | RRAM | SRAM | DRAM | MRAM | |
| Cell size | <4$F^2$ | 100$F^2$ | 6$F^2$ | 20$F^2$ | [37–41] F: feature size of lithography |
| Cell element | 1T(D)1R | 6T | 1T1C | 1(2)T1R | |
| Read /Write speed | ~300 ps | ~1 ns | ~10 ns | ~5 ns | |
| Endurance | $10^6$~$10^{12}$ | >$10^{16}$ | >$10^{16}$ | $10^{16}$ | |
| Retention | ~10yrs | Not specified | ~64ms | ~10yrs | |
| Capacity | 446 GB/$cm^2$ | Not specified | 46GB/$cm^2$ | Not specified | |
| Volatility | Non Volatile | Volatile | Volatile | Non Volatile | |
| Device Yield | 80-95% | Not Specified | 90% | Not Specified | |

| MLC Capability | 4bits/cell | 2bits/cell | 2bits/cell | 2bits/cell | |
|---|---|---|---|---|---|
| Scalability | Yes | No | No | Yes | |
| Program Energy/bit | 2pJ | Not specified | 2pJ | 120pJ | |

The recent research showed that there is a growing interest of research where natural materials are utilized as the active materials for the development of RS devices in order to enhance their performance with wide range of applications like the fabrication of implantable chips, electronic skin, biomedical diagnosis and therapy, artificial synapses[42–47]. The main advantage of using natural materials are their environment friendliness, biodegradability, pollution free, transient nature etc.. Even, the primary advantage of biological material based memory devices are (i) the abundant accessibility of useful material in our environment resulting in a low cost of this memory device, (ii) Bio- RRAM have amazing characteristics of self decomposition that can be sused for the applications of security and military purpose, (iii) Moreover, bio RRAM are compatible with living systems, which makes them an ideal candidate to be embedded into the body of a living being without any hazardous effect. Such implantation of bio-RRAM can be useful to study and even to control the behavior of animals[6]. Bio-RRAM also have high flexibility that is absent in most of the inorganic semiconductor based RRAM devices making them useful wearable electronics. Recent literature survey revealed that use of natural materials especially plants to realize electronics devices including the RS memory is growing enormously. In this paper, resistive switching phenomena using natural plant extract have been summarized highlighting the significant recent outcome. Such natural plant extracts have high potential towards the new generation environment friendly memory applications as well as other electronic devices.

**Plant Based RRAM device:**

There are numerous green materials which have been engaged in the fabrication of RRAM devices such as Aloe vera, lotus leaves, citrus leaves, dead leaves, maple leaves etc.[48–51]. Significant results related to natural materials based RRAM devices have been summarized in the following section of the manuscript.

**1. Aloe vera Based RRAM Device:**

Aloe vera is a succulent plant and based on its vital application, it is being cultivated around the world. Aloe Vera is composed of (i) a latex layer (protective layer), (ii) the epidermis (for the maintenance of water loss), and (iii) the parenchyma (for water storage), where the parenchyma is the innermost part of the Aloe vera leaf containing colourless and thin-walled tissues that further consist of three different structural components, namely (i) the cell wall (ii) mucilaginous gel and (iii) micro particles[6]. The main chemical compound present among these three components of parenchyma is carbohydrates. In addition to numerous other applications, it has been observed that Aloe vera based RS memory devices have high potential towards the next generation memory and may play an important role towards realization of green electronics[48, 52–55].

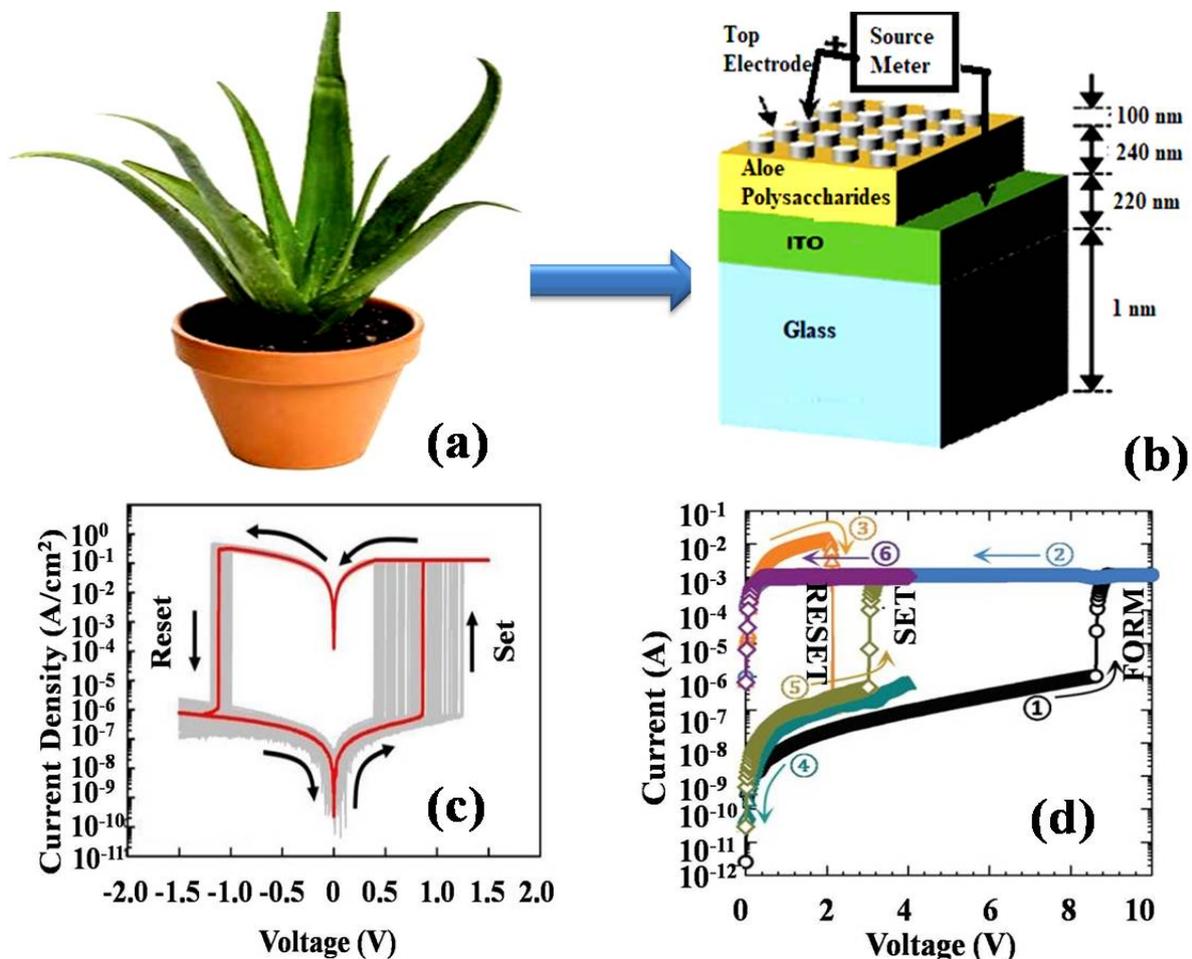

Fig 1: (a) Aloe vera plant (b) Device structure with Aloe vera as the active layer of RS. Both bipolar and unipolar resistive switching can be obtained within the Aloe vera polysaccharide based devices by varying top electrode. Reproduced with permission from[48]. (c) Experimental

(grey) and statistical (red) I-V curves corresponds to bipolar switching from Aloe vera based devices with Ag top electrode. Reproduced with permission from[55].(d) Typical I-V characteristics of polysaccharides device with AuTE showing unipolar switching. Reproduced with permission from[48].

Kuan Yew Cheong investigated that the typical current–voltage (I–V) characteristics of the polysaccharides based Aloe vera device having configuration of Cu/Aloe polysaccharides/ITO[48]. The device shows both bipolar and unipolar resistive switching based on the proper choice of top electrode. Polysaccharides are extracted from natural Aloe vera gel using facile alcohol precipitation. The devices are capable to deliver competitive performances, including a read memory window (ON/OFF ratio) as large as $\approx 10^7$ which allows them to be operational even in noisy environments. The switching parameters for each device are extracted from over 100 switching cycles. The resistive switching behaviors of such device can be explained for three distinct switching mechanisms attributed to various electronic, electrochemical, and thermo-chemical processes in the Aloe polysaccharides film[48]. It was also observed that electrode materials play crucial role towards the charge conduction mechanism in Aloe vera based switching device[48, 52]. Fig 1 shows the device structure and I-V behavior of Aloe vera based memory devices.

RRAM device based on Aloe vera with configuration of Al/Aloe vera/ITO on a nonflexible glass substrate has also been reported. This device exhibited bipolar memory switching with ON/OFF ratio $10^3$, retention time of more than $10^4$ s, and threshold voltages of $V_{SET}$ = −0.9 V and $V_{RESET}$= +3.1 V; however, the electrical endurance of this device was only 21. The conduction mechanism was deduced to be space charge limited conduction (SCLC) owing to the functional groups of de-esterified pectins and acemannan in the dried Aloe vera. The drying temperature of pure Aloe vera had a huge impact on the switching ratio. It was verified that only the device with Aloe vera dried at 50 °C showed a bipolar memory effect[53].

It has also been reported that Au nanoparticles (AuNPs)-loaded Aloe vera film when sandwiched between two electrodes shows resistive switching behavior that can be utilized as a RRAM for high-density information storage[54,55]. The memory device features a reasonably high ON/OFF ratio of $10^3$ at $V_{READ}$= 1.0 V and a wide read memory window of 4.0 V. Current density (at 5 V) recorded in the Aloe vera layer increases from $1.28 \times 10^{-8}$ to $5.09 \times 10^1$ A·cm$^{-2}$ as the temperature is increased from 40°C to 90°C. The ON state current remains relatively constant for $10^4$ s, while there is a noticeable increment in the OFF-state current after $10^2$s. It has also been observed that chemically doped Aloe vera flower extract show two distinct conducting states. Toggling between the states was possible through applied bias in a controlled manner leading towards memory application potential with distinct memory window[54]. Therefore, Aloe vera appears to be a promising candidate for electronic applications, as demonstrated by a series of electrical responses in the device with Aloe vera as the active component.

## 2. Citrus Based RRAM device

The natural material citrus is one of the most well known organic product fruit commodities in the world due to its nutritional values. Citrus is biodegradable, in addition to its low toxicity, biodegradability of citrus fruit makes it an ideal candidate for the fabrication of environment friendly electronics devices.

Kai-Wen Lin et al. have investigated RRAM device based on Citrus thin film with varying top electrodes viz. Al, Ti or Au and ITO as the bottom electrode[50]. It has been reported that resistive switching performances significantly depend on the top electrode, specifically; the reaction between top electrode material and the insulating layer material. Interestingly RRAM devices made with Al electrode showed an outstanding ON/OFF ratio $10^4$, which is significantly higher as compared to the devices with top electrode Ti or Au, Al and Ti electrode-based devices have relatively optimum performance in comparison to Au metal as top electrode with a very high memory window ratio and outstanding stability. Relatively high performances for the Al/Citrus/ITO device is attributed due the formation of native ($AlO_x$) layer and a formidable stability with rising temperature. The mechanism of such devices was deduced to be due to the formation of Carbon filament across the citrus layer in between two electrodes. Fig. 2 shows the memory characteristics of citrus based memory device along with the temperature dependent behavior.

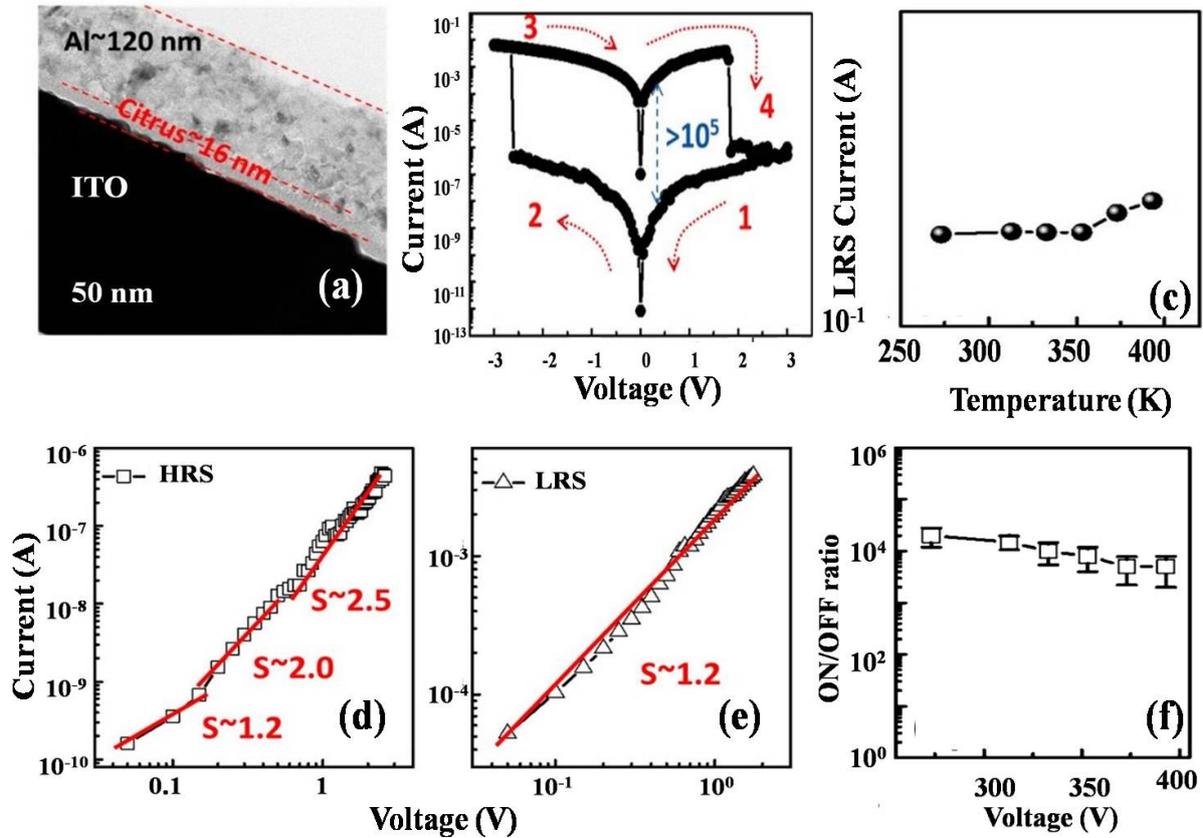

Fig. 2: (a) TEM image (cross-sectional view) of the Al/Citrus/ITO device. (b) Current–voltage (I–V) characteristics.(C) The ON/OFF ratio as a function of the temperature. (d) and (e) shows the double logarithmic plots of the high resistance state (HRS) and low resistance state (LRS) respectively of the memory device. (f) Temperature dependence of the LRS current of the citrus based memory device. Reproduced with permission from[50].

## 3. Lotus Based RRAM Device:

The lotus leaf is often used for medicinal herbs, cooking and viewing. Extracts from different parts of the lotus plant have been reported to show several biological activities, such as antioxidant, free radical scavenging, anti-inflammatory, and immuno-modulatory activities[49, 56, 57]. The main chemical compounds present in lotus leaf are carbohydrate, protein, fat, calcium and potassium. Lotus leaves are outstanding for being super hydrophobic. The outside of the lotus leaf has a gradually harsh structure. Lotus leaves, a prevalent aquatic herb serving as a dielectric layer, were used to design resistive switching devices. This may lead towards the realization of eco-friendly, low cost and degradable memory devices.

It has been reported that lotus leaves sandwiched between Ag and ITO electrodes showed bipolar resistive switching indicating potential application towards realizing organic bio-RRAM[49].The conduction mechanism across the devices was attributed due to the joint action of trap controlled SCLC and schottky emission.

Tengteng Li et al. has reported a pH controlled organic bio-memristor device using the lotus root as functional layer on a fixable substrate[57]. The device showed an excellent memristive memory performance at a pH 7.0 with stable memory behavior at least upto 100 consecutive cycles. Conduction mechanism was explained as both ohmic and SCLC model. This work opens up a new idea for the preparation of acid-base (pH) controlled memristor devices. Fig. 3 shows the characteristics of lotus based RS devices.

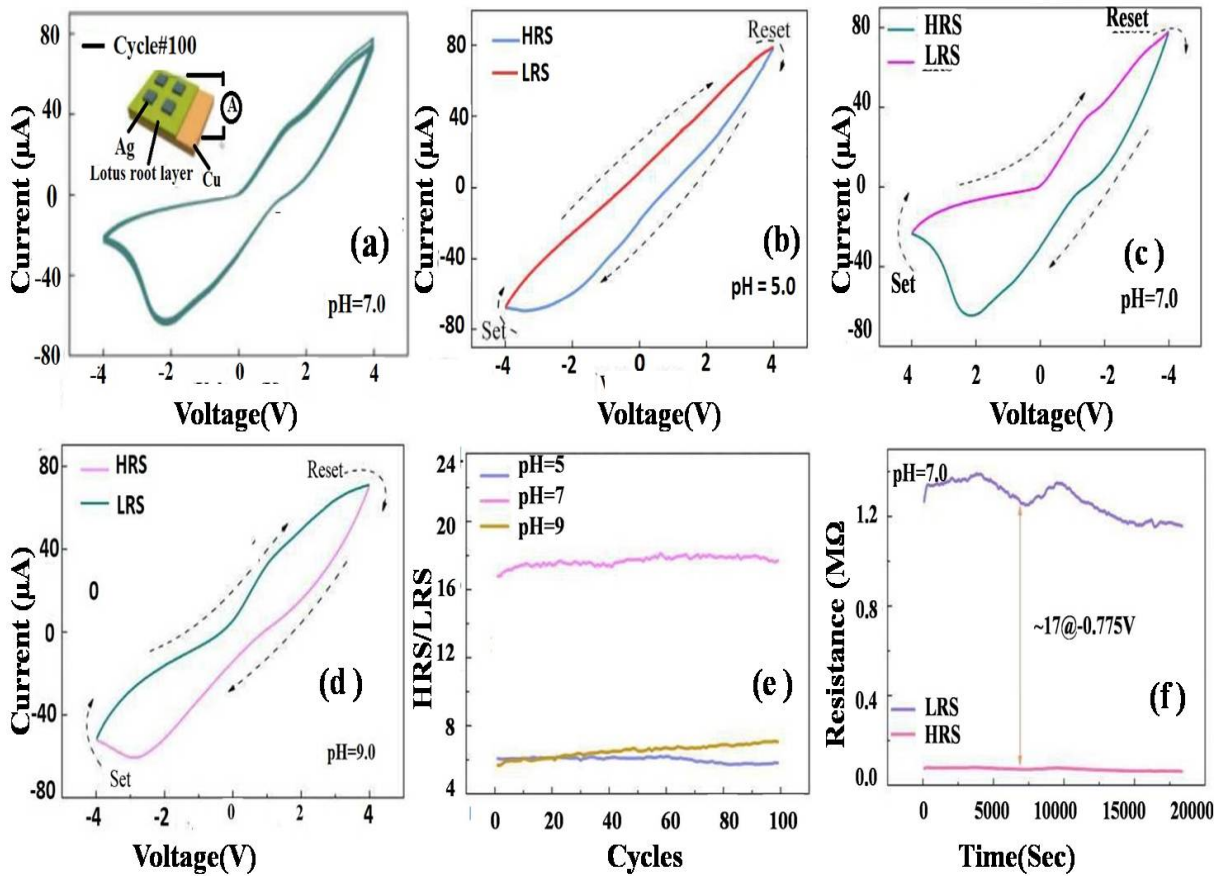

Fig. 3: (a) I-V curves with of Ag/Lotus root/Cu device under pH ~7.0. (b) I-V curve under pH ~5.0. (c) I-V curve under pH ~7.0. (d) I-V curve under pH ~9.0. (e) Different HRS/LRS

resistance ratios of the devices as a function of pH. (f) Retention behavior of lotus based device. Reproduced with permission from[57].

RRAM memory device based on Lotus leaves with configuration Ag/Leaves/Ti/PET structure on a flexible polyethylene terephtalate (PET) substrate has also been reported[56]. This device operates with $V_{SET}$ = 1 V and $V_{RESET}$ = −1.5 V and an electrical endurance sustained for more than 100 cycles without any substantial degradation in the memory window. Such bio- RRAM follows the conduction mechanisms of both ohmic behaviour (I ∝ V) along with SCLC (I ∝ $V^2$) at HRS, while in the LRS, the ohmic behaviour (I ∝ V) was followed. These results suggested that the lotus root-based organic bio-electronic device may be highly recommended for potential use in sustainable, low cost, and biodegradable memory applications. It also has a good potential to be used in environmentally benign flexible memory device.

## 4. Pristine Leaf Based RRAM device:

Adhikary et al. demonstrated that it is possible to design electro-chemical resistive switching device using freshly cut unprocessed leaves[58]. Freshly cut leaves of an indoor plant namely Golden Pothos (Epipremnum aureum), when directly connected with tungsten probes attached with a sourcemeter showed reliable resistive switching behavior at least upto 375 consecutive cycles between two resistance states with very high memory window. Such devices do not require any processing at all. Also such leaves are available in nature. Hence, this process is one of the most cost effective. Such devices may open a new path towards realizing transient electronics or in vivo integrated electronics within living creature for sensing purpose as working of the present device works based on the response of electrolytes within living cells.

## 5. Maple leaves Based RRAM device:

A multifunctional device having both memristive and capacitive effects have been reported using maple leaves (ML) and $TiO_2$ nanoparticle as the active layer[51]. It was possible to switch between capacitive behavior and memristive behavior upon application of suitable voltage. The device showed capacitive coupled memristive behavior and at low voltage and pure memristive behavior at high voltage. Such devices have been used to design low power passive filter demonstrating application towards reprogrammable analog circuit design. Fig 4 shows the preparation process and I-V curves of ML-TiO$_2$ (organic-inorganic) heterojunction device with structure Ag/TiO-ML/Al.

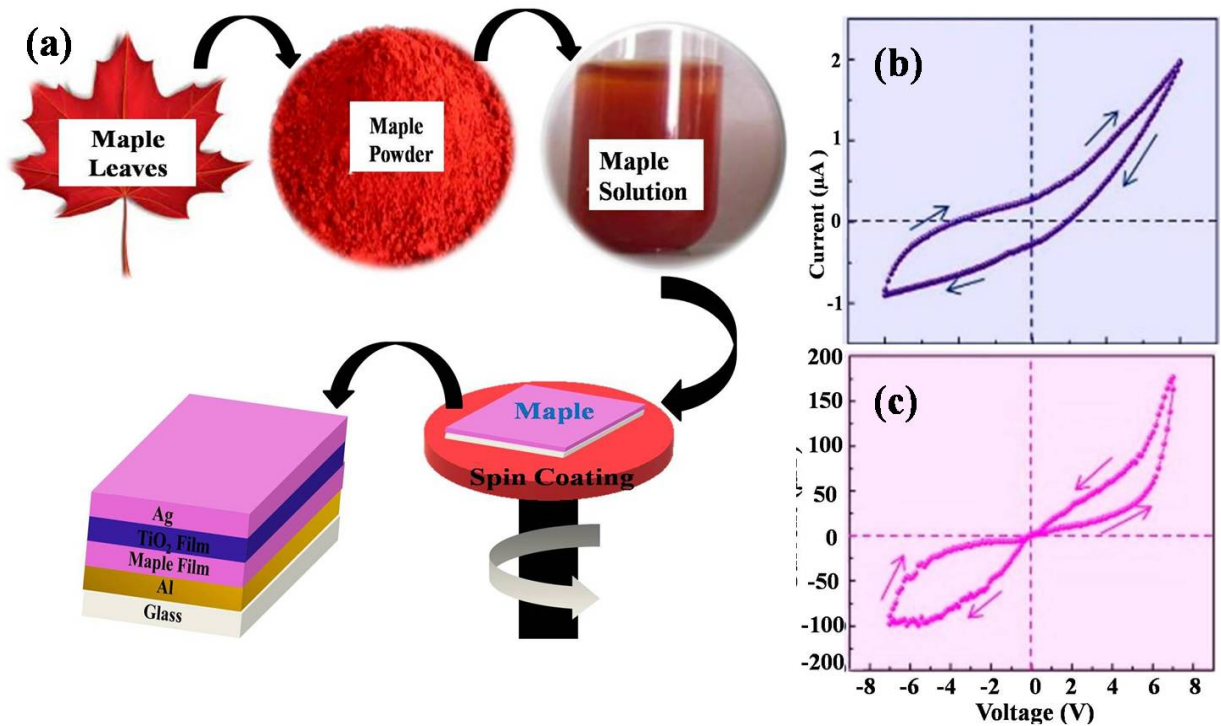

Fig. 4: (a) The preparation process of ML-TiO$_2$ (organic-inorganic) heterojunction device with structure Ag/TiO-ML/Al using maple leaves. (b) and (c) shows the capacitive coupled memristive behavior at low voltage and pure memristive behavior at high voltage. Reproduced with permission from[51].

## 6. Lophatherum Gracile Brongn based memory device:

Bio-memristor employing natural Lophatherum Gracile Brongn (LGB) as the active layer of the device has been reported[59]. Such device showed capacitive–coupled memristive behavior. Interestingly the capacitive coupled memristive behavior was found to be regulated by doping of Ag nanoparticle within the active layer (LGB) of the device. Ag ion transfer as well as Ag filament formation under the applied bias were the key behind the observed capacitive coupled memristive behavior in such devices.

**7. Lichen plant based memory device:**

Organic compound extracted from lichen plant has also been used to implement RS memory with multilevel operations and multi-state behaviors with inherent application potential towards neural network[60]. The device showed unique capacitive effect, negative differential resistance (NDR) along with multistage resistances. A physical mechanism, involving the water molecule reaction with oxygen vacancies, migration of ions, structural changes in $(C_7H_7O_4N)_n$, and the formation of emf is proposed to explain the origin of the multi-state RS behavior. Such device showed transition from capacitive behavior to resistive switching upon change of bottom electrode from ITO to Ti. Fig. 5 shows the structure and characteristics of corresponding RS devices.

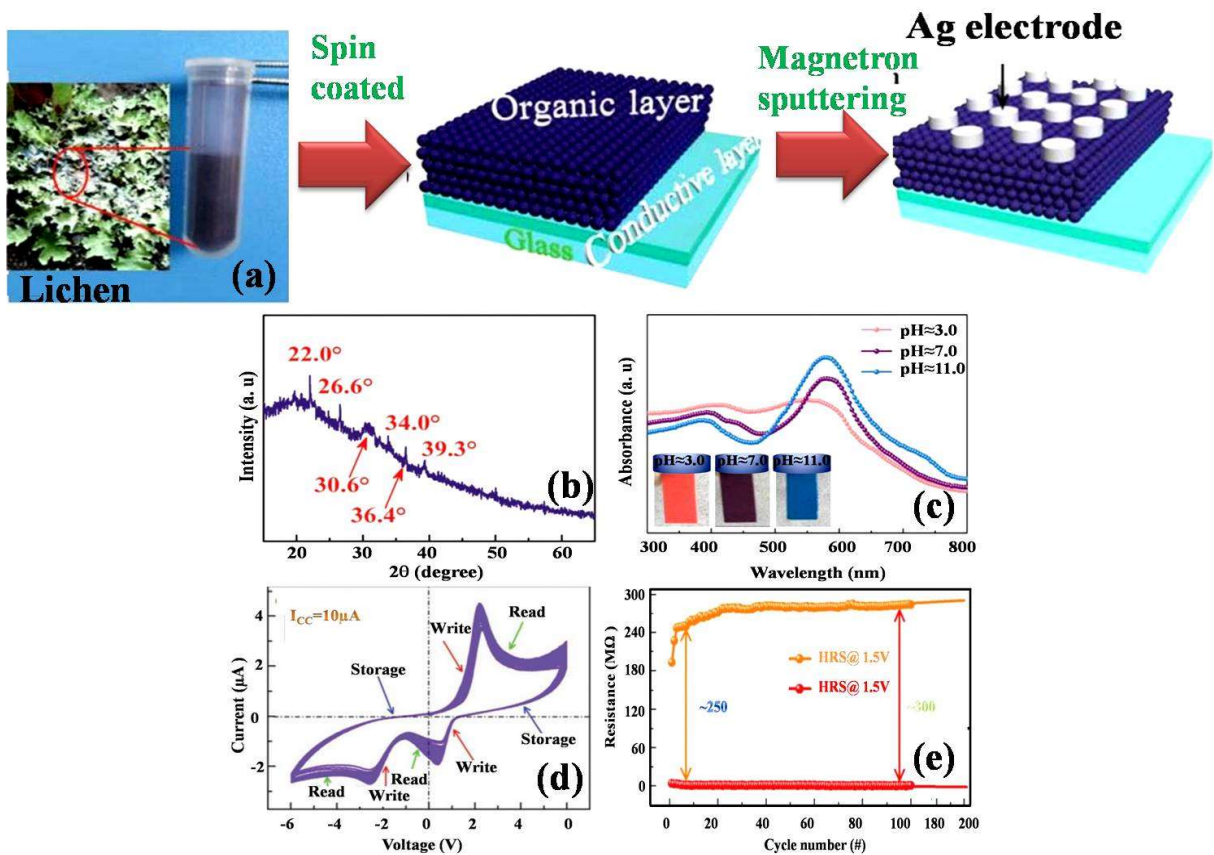

Fig 5: (a) Preparation process of lichen extract based memory device. (b) The X-ray diffraction (XRD) spectrum of the as –prepared $(C_7H_7O_4N)_n$ powder as isolated from lichen extract. (c) The UV-Vis spectra of the $(C_7H_7O_4N)_n$ powder at different pH conditions. (d) I–V characteristic curves of continuous 100 cycles for the Ag/$(C_7H_7O_4N)_n$/FTO device under a compliance current of 10.0 mA. (e) The HRS/LRS resistance ratio as a function of cycle number under a positive bias voltage of 1.5 V. Reproduced with permission from[60].

## 8. Natural collagen based RRAM device:

Collagen, extracted from pigskin has been utilized as the intermediate insulating layer for constructing two bio-memristive memory devices with Ag/Bio-film/ITO and Ag/Bio-film/Ti structures[61]. Both the device showed RRAM behavior with very good memory window. Switching mechanism in such device has been explained due to the charge filling of the defect states within the active layer. Biodegradability and environmental friendliness of the device indicates the application of such devices towards bio-electronics.

**9. Rose petal based WORM memory device:**

Rahman et. al. explored the rose petal based WORM memory device[43]. The device demonstrates non volatile resistive switching with excellent memory window, data retention, read endurance and device stability. Space charge limited conduction and ohmic conduction were involved in the conduction mechanism of such device. The results indicated that biodegradable, nontoxic natural rose petals opens up a new avenue towards the application in the non-volatile memory device.

**10. Potato Tuber Based RRAM device:**

RRAM device based on Potato tuber has also been reported[62]. The device exhibits bipolar switching behavior with rapid conversion and great reproducibility with very high ON/OFF ratio and long retention time. The Ag-electrode-based sample has showed optimum performance in comparison to other metals with a very high HRS/LRS resistance ratio and outstanding stability. Therefore, the Ag electrode plays a significant role in the conduction mechanism of this bio-RRAM. The conduction mechanism of this device was deduced to be as the joint action of SCLC effect and schottky emission.

**Fruit Based RRAM devices**

Fruit has tremendous potential to rectify several activities such as health hazardous medicinal use. Even it is biodegradable, non toxic, environment friendly substance that can also be used for electronic purposes. Several reports have been carried out in fruits such as orange, banana, apple etc to demonstrate the RS memory behavior for future memory application as elaborated in the following section[63–66].

**1. Orange Peel Based RRAM device:**

Orange peel is an un-useful by product from orange having tremendous impact in the medicinal properties. In accordance to that it is nontoxic, pollution-free, renewable, environment-friendly, and naturally decomposable substance. Recently orange peel has been used as the active layer of the RS memory devices, which showed non-volatile bipolar memory behavior[63].

The device exhibits bipolar switching behavior with rapid conversion and great reproducibility. Such device showed good memory characteristics such as high ON/OFF ratio and long retention time. The HRS/LRS ratio in term of device temperature indicates good memory behavior at room temperature. The conduction mechanism in the HRS of such device can follow the SCLC due to the defect arose in the active layer. The electronic devices using such nontoxic natural biomaterials may have great potential in wearable equipment, medical facilities and implanted device etc.

## 2. Banana Peel Based RRAM device:

Banana is a very common and largely consumed healthy fruit across the globe. Banana is generally the best choice for vitamin supplements. Annual output is about 1,200 tons in China. However, the usage of banana peel is scarce, currently used as fertilizer or discarded in most of the cases. It has been observed that the banana peel can be used as a raw material for assembly of memory devices, having numerous advantages, such as easy to draw, low cost, environmental friendly and biodegradable.

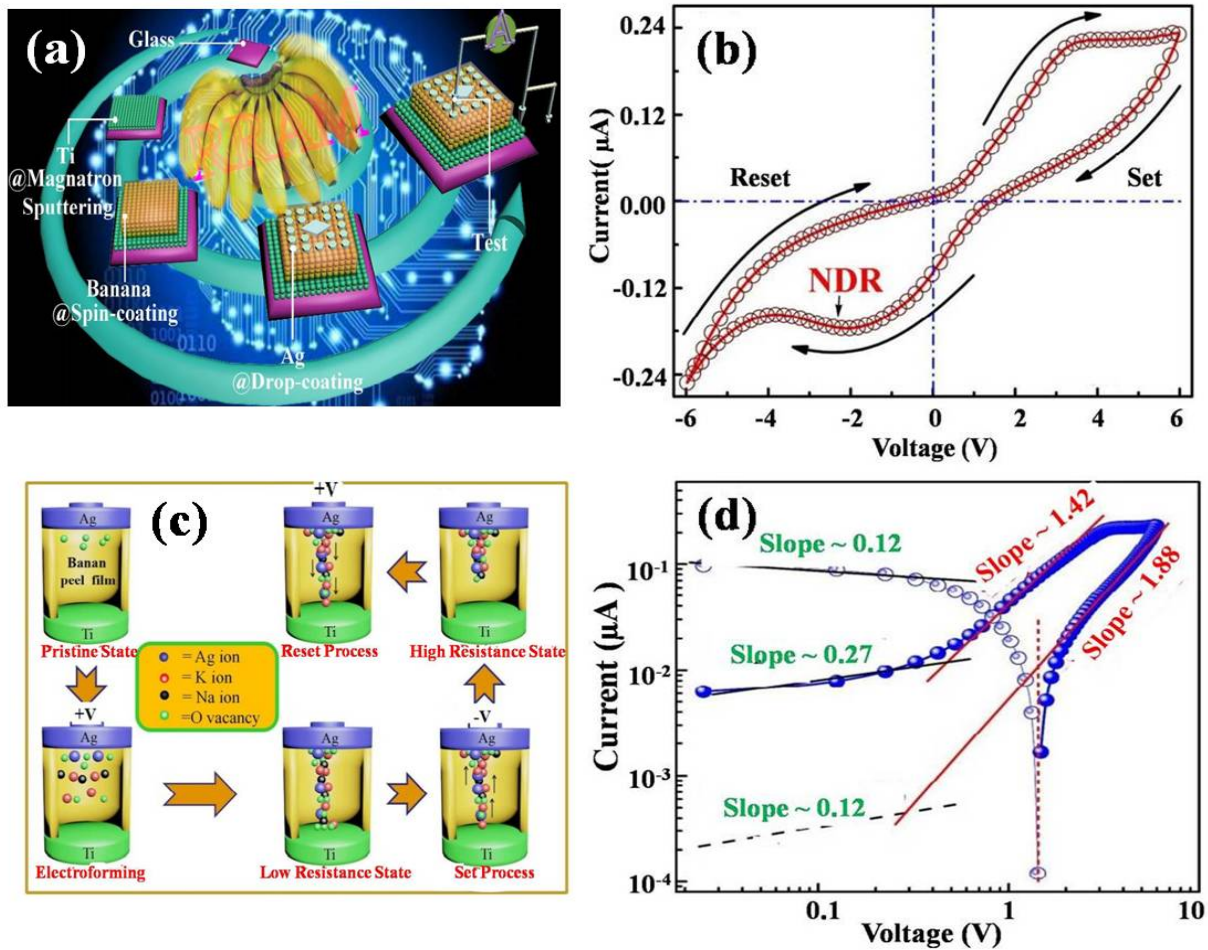

Fig 6: (a) Schematic of the device preparation process. (b) Hysteresis loops of the current-voltage characteristics for Ag/Banana peel/Ti structure. (c) Schematic diagram of the ion transfer process. (d) The positive region of I-V curves in double log scale, the solid spheres are experimental data and the straight lines are the linear fitting. Reproduced with permission from[64].

Liang et al. explored metal ions redox induced nonvolatile RS memory devices with banana peel as the active layer material[64]. Device structure and I-V curve of such devices are shown in Fig. 6(a). The device holds large switching resistance ratio and long retention characteristics, excellent cycling performance at ambient temperature. The conduction mechanism in such devices has been explained as the classical trap controlled SCLC. In addition, the co-existence of memristor effect, capacitance effect and NDR phenomenon is noticed in such device.

**3. Apple pectin based resistive memory:**

Yu-Chi Chang et. al reported the resistive switching behavior of apple pectin (AP) based memory devices with device structure Al/AP/ITO[66]. Pectins are abundant in fruits and may also be extracted from cell walls similar to cellulose extraction. The properties of pectin include gelatification, thickening, and stabilization, enabling it to be one of the most commonly used plant polysaccharides in the food and medical industries. It has been demonstrated that it was possible to control the filaments formation types either metal or carbon by controlling the electrode (Al) and apple pectin interface. The AP memory device showed a promising ON/OFF ratio $\sim 10^7$ with uniform electrical distribution and stable retention behavior. Exploiting the dependence of the different characteristics of the Al/AP interface on resistive switching may enable the designs for future smart bioelectronics.

## 4. Lime peel based RRAM device:

It is well known that leakage of current in oxide layers is the main concern for higher speed and denser RRAM. Defect engineering played main role in meeting this challenge by doping or controlling the defects or active sites. Device based on lime peel extract (LPE) with configuration Ag/$TiO_2$-LPE/FTO has been studied manipulating the defect within the active layer[65]. This device was developed in order to resolve the sneak path current issue with the utilization of active interstitial defects or active sites followed by the electrochemical metallization (ECM) mechanism. Transition metal oxides (TMOs) are capable of gaining greater attention as it is thermally stable, has excellent optical and dielectric characteristics with a high dielectric constant value, additionally biocompatible and non toxic as compared to other metal oxide elements[65]. Moreover, $TiO_2$ NPs are having antibacterial effect so with the incorporation of this NP in flavone-enriched LPE makes $TiO_2$ a superior antibacterial agent towards different infectious pathogens. The designed memory showed excellent endurance, a high ON/OFF ratio without any electrical degradation. The retention test indicates that the LPE memory cell retain its HRS and LRS states over >$10^5$ s with no apparent electrical degradation having excellent ON/OFF ratio $10^5$. The classical SCLC and electrochemical metallization (ECM) are two conduction mechanisms for this bipolar resistive switching behavior.

## 5. Grape seed based RRAM device:

Sun et al. reported the battery-like resistive switching behavior in self resulting bio-memristive using $C_{15}H_{11}O_6$, extracted from grape seeds[67]. The biomaterial anthocyanin/graphene composite was chosen for the switching layer in the ReRAM device in order to establish the ionic behaviors under electric field. Due to the incorporation of graphene and ionic liquid decoration, a high memory window with excellent operation of 100 switching cycles has been achieved. Self-selecting bio-memristor can potentially be applied in the potentiation and depression of artificial synapses which enables the neuromorphic computing. Such device may be used for the application of electronic skin and artificial intelligence. Such self resulting memristor can also be very useful to develop compact and concise integrated circuits.

**Miscellaneous:**

In addition to the above, there exists some more reports of resistive switching devices using various other natural materials like garlic, honey milk etc.[68–71]. Such results have been summarized in the following sections.

**1. Garlic based RRAM device:**

Garlic, among the oldest cultivated plants, is used both as a food and for medicinal application. It contains antioxidants that support the body's protective mechanisms against oxidative damages. Additionally, it is also used as a medicine for the treatment of high blood pressure. Moreover, it is environment friendly, sustainable, pollution-free, and biodegradable. Interestingly, it has been observed that garlic extract has been used for designing memristive device[69].

Shuangsuo et. al. designed a new type of memristive memory device with edible garlic as the active layer for analog neuromorphic sensor applications[69]. This device can be precisely modulated by varying the pH that can mimic neuron function and can be observed as a computing resource exploiting neurons in biological systems. The pH-controlled steps of signal processing, represents the application in hardware implementation for neuromorphic formalities. Memristive device based on garlic exhibits non-volatile memristive memory behaviour, at very low operation voltage (~1.0 V), long-time data retention (~30000 s), and with enough stability. The device shows durable and reliable memristor behaviour and can be maintained for more than 200 cycles. These data indicated that garlic based memristor may have potential application for simulating synapses and/or wearable electronic devices.

## 2. Honey Based RRAM devices:

Honey is a natural sweet product containing more than 180 constituents mainly sugar. The role of honey has been acknowledged in scientific literature and there is persuading evidence in support of its antioxidant and antibacterial nature, cough prevention, fertility and wound healing properties. Aleksey A. Sivkov et al. demonstrated RS behaviour using honey having device configuration of Cu/Honey/ $Cu_xO$ on a glass substrate[70]. The Current-voltage measurements showed that the device exhibited bipolar switching characteristics suitable for RRAM application with a very high memory window of the order of $10^7$.

## 3. Mushroom based RRAM device:

Mushrooms are rich, low calorie source of fiber, protein, and antioxidants. RS behavior with edible mushroom extract as an intermediate insulating material has also been reported[71].

Metals such as Al, Cu, Ag, and Ti were chosen as the bottom electrode in order to explore the physical mechanisms in such devices. It was found that the redox of hydroxyl-assisted Ag filaments could be easily formed on an inactive metal bottom electrode through the mushroom film due to redox reaction. Maximum current is set at 1.0 mA in order to refrain from permanent damage to the mushroom film that figure out the low power consumption of the device. The devices exhibited reproducible bipolar resistive switching memory behaviour with very good memory window for at least 100 consecutive cycles when Ti was used as the bottom electrode. Possible underlying switching mechanism of the memory was the redox of hydroxyl-assisted Ag-rich conduction. Classical SCLC was also involved in the conduction process. This study indicates that mushroom is a promising material for bio inspired and biocompatible bio-RRAM. It has been believed that this flexible and biocompatible memory device comes up with an utmost step in the development of implantable and bioelectronics. Fig 7 shows the I-V characteristics of the mushroom based RS device.

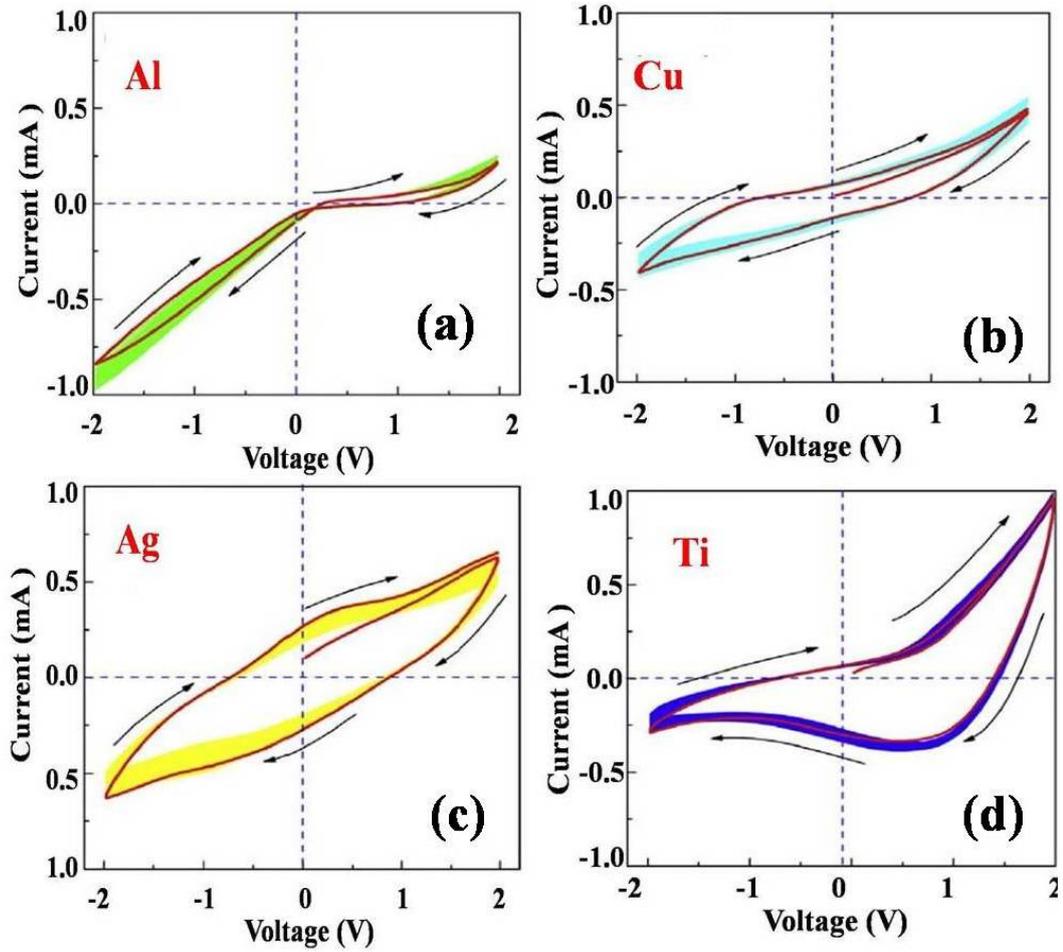

Fig 7: I-V characteristics of the Ag/mushroom/metal (Al, Cu, Ag, Ti) device. It was possible to obtain memory behavior starting from hysteresis behavior upon changing the bottom electrode from Al to Ti. Reproduced with permission from[71].

## 5. Milk based RRAM device:

Lactose is the abundant material that exists in the milk. Environment friendly, disposable and transient RS memory using α-lactose derived from milk has been demonstrated[68]. Device structure and characteristics are shown in Fig. 8. Electrochemical formation/rupture of the Ag metallic filaments across α-lactose layer is believed to be responsible for the switching behaviour of the memory devices. These device exhibits operation voltages of 1V, data retention 5000s and switching cycles 100 with the ability of the multilevel storage. α-Lactose film can be dissolved completely in demonized water within 3s, showing its physically transient nature and biodegradable characteristics. Such devices may be well adjusted for fabrication of skin compatible, digestible bio-electronic devices.

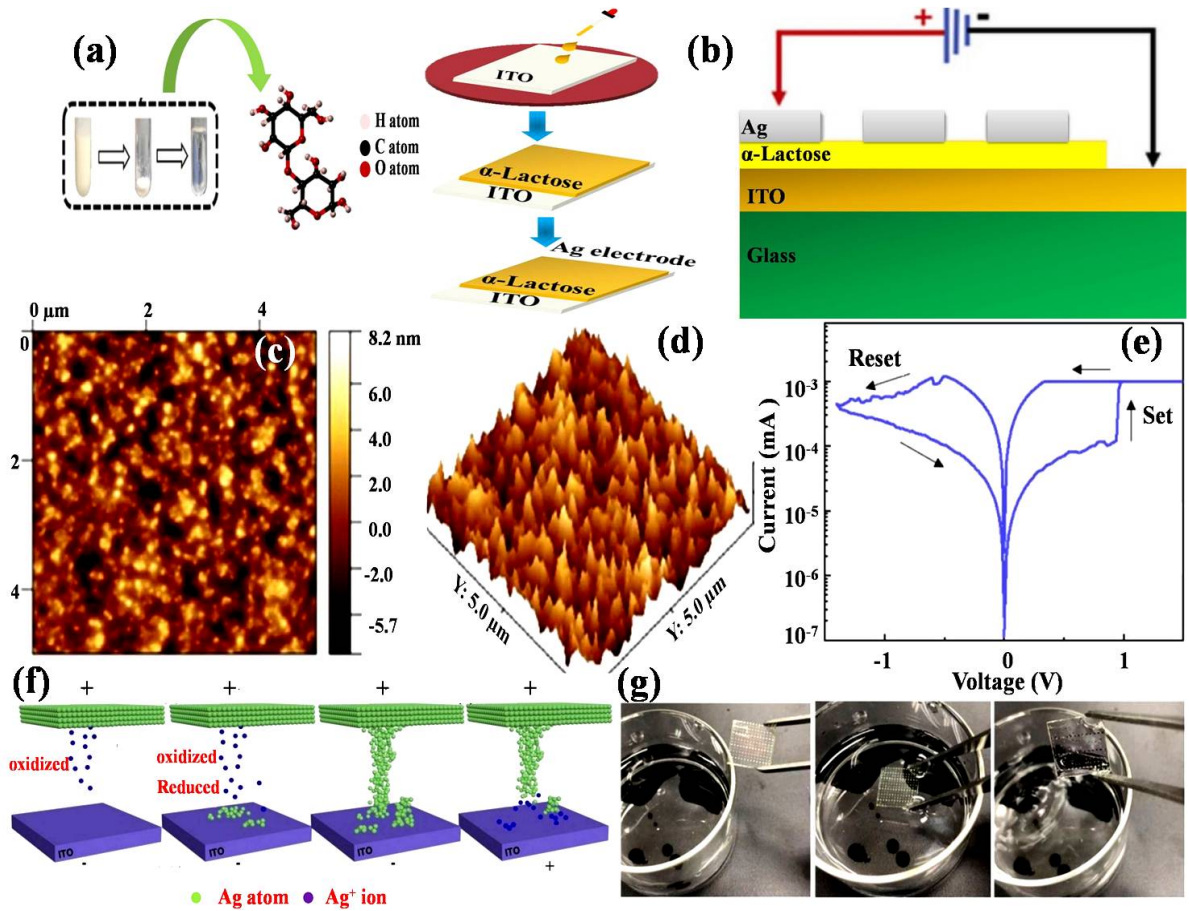

Fig. 8.(a) The specific manufacturing process of Ag/α-lactose/ITO memory cell. (b) Structure diagram of the Ag/α-lactose/ITO device. (c) The Atomic Force Microscope (AFM) topography image of the α-lactose film. Right: Three-dimensional AFM image corresponding to the left. (e) Typical bipolar resistive switching I–V curves of the Ag/α-lactose/ITO device. (f) The schematic diagrams for formation rapture of the Ag filament. (g) Dissolution of α-lactose thin film with deionized water at room temperature using Ag/α-lactose/ITO configuration. Ag electrode which is insoluble in water has left the ITO surface and floated in the water. Reproduced with permission from[68].

## 6. Sweet potato peel based RRAM device:

Sweet potatoes is a vital crop and the non starch polysaccharides in the sweet potato peel (SPP) mainly originates from the complex carbohydrates of plant cell, wall. It has been observed that SPP may be a good complement for developing organic memristor[72].

The functional layer based on sweet potato peel was used to design a memristive device with Ag/SPP/F-doped $SnO_2$ (FTO) structure for non-volatile applications. Such device exhibits a stable HRS/LRS resistance ratio and longer retention time at room temperature. The non zero crossing I-V hysteresis behavior has also been shown due to the charge transfer at the Ag/SPP and SPP/FTO interfaces. SCLC was also involved in the conduction process within such devices. Rich I-V characteristics, excellent non volatile behavior of SPP based device opens up a goal towards the advancement of electronic devices with sustainable technology.

## 7. Natural melanin based RRAM device:

Memristive device with melanin as the active layer with remarkably good endurance and retention characteristics has also been demonstrated[73]. Natural melanin has been isolated from bacterium Aeromonas sp. SNS. The device was electroforming-free and required very low switching voltages (0.6 V) having very low power dissipation during the resistive switching process. The melanin based memristive device displays rich nonlinear behavior and exhibits negative differential resistance effect. The formation of conductive filament due to the SCLC were responsible for the observed bipolar resistive switching effect.

**Table 2: List of natural materials used as active materials for RS devices along with switching/memory parameters:**

| Natural materials | Device structure | On/off ratio | Endurance | Memory type | Retention (sec) | References |
|---|---|---|---|---|---|---|
| Aloe Vera | Cu/aloe polysaccharides/ITO | $10^7$ | 100 cycles | RRAM | Not specified | [48] |
| | Al/Aloe vera/ITO | $10^3$ | 21 cycles | RRAM | $>10^4$ | [53] |
| | (Al, Ag))/Aloe Vera/ITO | $10^3$ | 100 cycles | RRAM | $10^4$ | [52] |
| | Al/Aloe vera+ AuNPs /ITO | $10^3$ | Not specified | RRAM | $10^4$ | [54] |
| Citrus | Al, Ti, Au /citrus/ITO | $10^4$ | 50 cycles | RRAM | $10^4$ | [50] |
| lotus leaves | Ag/Lotus leaves/ITO | About 40 | 100 cycles | Bio-RRAM | $10^3$ | [49] |
| lotus root | Ag/lotus root/Cu | Not specified | 100 cycles | RRAM | $>10^4$ | [57] |
| Lotus dead leaves | Ag/Leaves/Ti/PET | About 30 | 100 cycles | RRAM | $10^3$ | [56] |
| Pristine leaves | Leaf with probes as electrodes | 73 | at least 375 cycles | RRAM | Not specified | [58] |
| Mapple leaf | Ag/TiO$_2$-ML/Al | Not specified | 30−100 cycles | Capacitive-coupled memristive behavior | Not specified | [51] |
| Lophatherum gracile Brongn | Ag/LGB/fluorine-doped tin oxide (FTO) | 23.4 | 150 cycles | Bio-memristor | Not specified | [59] |
| lichen plant | Ag/(C$_7$H$_7$O$_4$N)$_n$/ F-doped SnO$_2$ (FTO) | $>10^2$ | 100 cycles | pH controlled RRAM | $10^3$ | [60] |
| Natural Collagen | Ag/natural collagen /ITO | 100 | 75 cycles | RRAM | $10^3$ | [61] |
| Rose petal | Au/Rose/ITO | $>10^2$ | 10 cycles | WORM | 7200 | [43] |
| Potato tuber | Ag/AgCl | $10^3$ | Not specified | Bio RRAM | Not specified | [62] |
| Orange peel | Ag/Orange peel/FTO | $10^3$ | >100 cycles | bipolar switching | 500 | [63] |
| Banana peel | Ag/banana peel/Ti | ~60 | Not specified | Bio-RRAM | $>10^4$ | [64] |
| Apple | Al/AP/ITO | $10^7$ | 50 cycles | Biomemory | $>10^7$ | [66] |
| Lime peel | Ag/TiO$_2$- LPE/FTO | $10^5$ | $1.5 \times 10^3$ cycles | Bio-RRAM | $>10^5$ | [65] |

| Grape seed | Mo/$C_{15}H_{11}O_6$–graphene/Mo | Not specified | 100 cycles | RRAM | Not specified | [67] |
| --- | --- | --- | --- | --- | --- | --- |
| Garlic | Ag/garlic/fluorine-doped SnO2 (FTO) | 25 | >200 cycles. | pH-Modulated memristive behavior | >$10^4$ | [69] |
| Honey | Cu/Honey/$Cu_xO$ | $10^7$ | Not specified | ReRAM | $10^4$ | [70] |
| Mushroom | Ag/mushroom/ (Al, Cu, Ag, and Ti) | 1.2,5,7,10 | 100 cycles | Bio-RRAM | Not specified | [71] |
| Milk | Ag/ α-lactose/ITO | Not specified | 100 cycles | RRAM | 2000 | [68] |
| Sweet potato peel | Ag/SPP/FTO | Not specified | 200 cycles | Memrister | $10^4$ | [72] |
| Natural melanin | Ag/melanin/SS | 10x | >$10^4$ cycles | RRAM | $10^3$ | [73] |

**Conclusion and future outlook:**

In this paper, we have summarized the recent results related to bio-based resistive switching memory devices employing natural plant extract. It has been observed that a good number of research works are going on across the globe related to this area. Resistive switching devices, specially using plant materials are very promising as these are most abundant renewable material on the earth. Several significant results have already been reported like multifunctional device having memristive and capacitive effects, resistive switching using apple pectin, pH controlled garlic based memristive device for neuromorphic application etc. In some cases, pristine leave based device showed reliable memory behavior even beyond 375 consecutive cycles of operations, apple based device was able to retain the stored information atleast upto $10^7$sec with high memory window $10^7$ etc. As a whole, it has been noticed that a significant amount of work is going on where natural materials especially plant extract have been used to realize RS memory device. Such devices have tremendous potential to provide sustainable solution towards memory technology due to their quick programmability, high memory window, cyclibility, low cost, low power consumption, easy fabrication, down scalability, flexibility, high data retention and stability etc.

Despite of several interesting developments, still there are certain issues which need to be addressed for real practical applications.

(i) Natural material/ biomaterials decompose very fast. Therefore mechanical stability as well as life span of electronic device using natural material need to be improved.

(ii) Plant extract contains numerous compounds, therefore, purification and quantification is a complex process and big challenge.

(iii) In vivo conditions in the human body are completely different from that in vitro in a lab. Therefore, further detail investigations are required for real transient and implantable memory/sensing, neuromorphic computation and other biomedical applications aiming to lower the gap between in vivo and in vitro operations.

(iv) In general, accurate deep theoretical knowledge explaining the conduction mechanism in case of RS devices is yet to be developed. Therefore, co- relation between theory and experimental observations are yet lacking.

(v) One of the important key factor is the switching speed between two memory states (ON/OFF) for optimum performances and natural material based RS devices do not possess fast switching speed. Therefore, for real practical applications, this need to be taken care of.

A multidisciplinary outlook and collaborative efforts from different branches of science like- physics, chemistry, engineering and biology may surely able to resolve these issues. In this regard it is high time to make a great stride in these emerging areas of research. This may provide a conceptual understanding with wide scope of real practical applications. At the end it can be said that recent extensive research efforts towards natural material based RS memory indicate the probability of commercialization/technological applications in near future owing to their remarkable advantages like transparency, biodegradability, environmental friendliness , tunability, flexibility as well as robustness. This will open a new dimension for wearable and implantable memory/technology.